\documentclass[12pt]{article}

\usepackage[margin=1in]{geometry}
\usepackage{amsmath,amssymb}
\usepackage{graphicx}
\usepackage{booktabs}
\usepackage{natbib}
\usepackage{hyperref}
\usepackage{xcolor}
\usepackage{float}
\usepackage{caption}
\usepackage{setspace}
\usepackage[T1]{fontenc}
\usepackage[utf8]{inputenc}
\usepackage{lmodern}  
\usepackage{array}
\usepackage{longtable}
\usepackage{threeparttable}

\hypersetup{
  colorlinks=true,
  linkcolor=blue!60!black,
  citecolor=blue!60!black,
  urlcolor=blue!60!black
}

\title{Scaffolding Human--AI Collaboration:\\A Field Experiment on Behavioral Protocols and Cognitive Reframing}

\author{
  Alex Farach\thanks{Corresponding author. Microsoft Corporation. Email: \texttt{alexfarach@microsoft.com}} \and
  Alexia Cambon \and
  Lev Tankelevitch \and
  Connie Hsueh \and
  Rebecca Janssen \\[6pt]
  \normalsize Microsoft Corporation
}

\date{March 2026}

\begin{document}

\maketitle

\begin{abstract}
Organizations have widely deployed generative AI tools, yet productivity gains remain uneven, suggesting that \emph{how} people use AI matters as much as whether they have access. We conducted a field experiment with 388 employees at a Fortune 500 retailer to test two scaffolding interventions for human--AI collaboration. All participants had access to the same AI tool; we varied only the structure surrounding its use. A behavioral scaffolding intervention (a structured protocol requiring joint AI use within pairs) was associated with lower document quality relative to unstructured use and substantially lower document production. A cognitive scaffolding intervention (partnership training that reframed AI as a thought partner) was associated with higher individual document quality at the top of the distribution. Treatment participants also showed greater positive belief change across the session, though sensitivity analyses suggest this likely reflects recovery from carry-over effects rather than genuine training-induced shifts. Both findings are subject to design limitations including an AM/PM session confound, differential attrition, and LLM grading sensitivity to document length.

\medskip
\noindent\textbf{Keywords:} generative AI, human--AI collaboration, organizational adoption, field experiment, scaffolding
\end{abstract}

\newpage

\section{Introduction}
\label{sec:intro}

\subsection{The AI Integration Challenge}

Generative AI tools are now available to most knowledge workers, yet organizations report uneven adoption and uncertain returns on AI investments \citep{brynjolfsson2023}. Unlike previous enterprise technology where access was the primary barrier, the bottleneck appears to have shifted from technical availability to human integration, meaning how people actually incorporate AI into their work processes.

Early field evidence reveals substantial heterogeneity in productivity effects. \citet{noy2023} find that AI assistance improves writing task completion time by 40\% on average, but with considerable variance across workers. \citet{dellacqua2023} demonstrate that AI can improve consultant performance within the ``jagged technological frontier'' of tasks where AI is capable, but harms performance when applied beyond AI's competence. Workers struggle to identify which tasks fall within this frontier, suggesting that effective AI use requires not just access but judgment about when and how to deploy AI assistance.

This suggests a central question for organizational research: if access is no longer the bottleneck, what interventions help people use AI more effectively?

\subsection{Two Approaches to Scaffolding AI Use}

We distinguish two forms of intervention (behavioral scaffolding and cognitive scaffolding) drawing on the educational concept of scaffolding as temporary support that enables performance beyond independent capability \citep{vygotsky1978,wood1976}.

\textbf{Behavioral scaffolding} refers to explicit protocols that structure how humans interact with AI systems. Drawing on socio-technical systems theory \citep{trist1951,orlikowski1992}, we reason that without explicit guidance, users may default to familiar, low-risk interaction patterns, treating AI as an advanced search engine rather than integrating it into collaborative work \citep{jasperson2005}. As algorithms have become more embedded in organizational workflows, they have reshaped expertise boundaries and introduced new forms of control over worker behavior \citep{faraj2018,kellogg2020}. Behavioral scaffolds create structural conditions for more sophisticated interaction, particularly in team settings where AI mediates human-to-human collaboration.

\textbf{Cognitive scaffolding} refers to interventions that reshape users' mental models of AI. Building on research on technology frames \citep{orlikowski1994}, we propose that users' implicit beliefs about AI (whether it is a ``tool'' to be operated or a ``collaborator'' to engage with) may shape their interaction patterns. \citet{cadario2021} demonstrate across five experiments that brief cognitive reframing interventions can shift AI utilization, and \citet{lebovitz2022} find that only professionals who developed ``engaged augmentation'' practices (actively interrogating AI outputs rather than passively accepting or rejecting them) realized the benefits of AI integration. Cognitive scaffolds aim to shift users from a transactional mental model (query-response, one-shot) to a dialogic one (iterative, conversational). This distinction parallels the theoretical tension between structural and interpretive approaches to technology adoption \citep{desanctis1994,weick1990}.

\subsection{Human--AI Collaboration Beyond the Individual}

Most research on AI productivity focuses on individual human--AI interaction. Yet knowledge work is fundamentally collaborative, with outcomes depending on collective intelligence---a group's ability to solve problems through distributed cognition \citep{woolley2025}. A recent meta-analysis of 106 studies finds that human--AI combinations actually perform \emph{worse} than the best of humans or AI alone on average, though content creation tasks are a notable exception where combinations show gains \citep{vaccaro2024}. \citet{dellacqua2025}, in a field experiment at Procter \& Gamble, found that teams working with AI achieved the highest performance, suggesting AI and human collaboration can be additive. However, \citet{schmutz2024} find that adding AI to teams often reduces coordination, communication, and trust, which implies that simply inserting AI into collaborative workflows without addressing team dynamics may undermine performance. \citet{doshi2024} find that AI assistance improves individual creative output but homogenizes collective output, reducing the diversity of ideas across a group.

These findings motivate our behavioral scaffolding intervention: a structured protocol designed to maintain human-to-human connection while integrating AI as a shared tool, rather than allowing each team member to use AI individually and then merge outputs.

\subsection{Present Study}

We report findings from a field experiment with 388 employees at Gap Inc., a Fortune 500 retailer. The experiment tested two interventions within a single day:

\begin{itemize}
  \item \textbf{Task A} (pair-level): A structured ``Create-Out-Loud'' collaborative protocol requiring joint AI use, compared against naturalistic AI use where pairs worked however they chose.
  \item \textbf{Task B} (individual-level): Partnership training reframing AI as a ``thought partner,'' compared against standard technical training on AI features.
\end{itemize}

\begin{figure}[htbp]
  \centering
  \includegraphics[width=\textwidth]{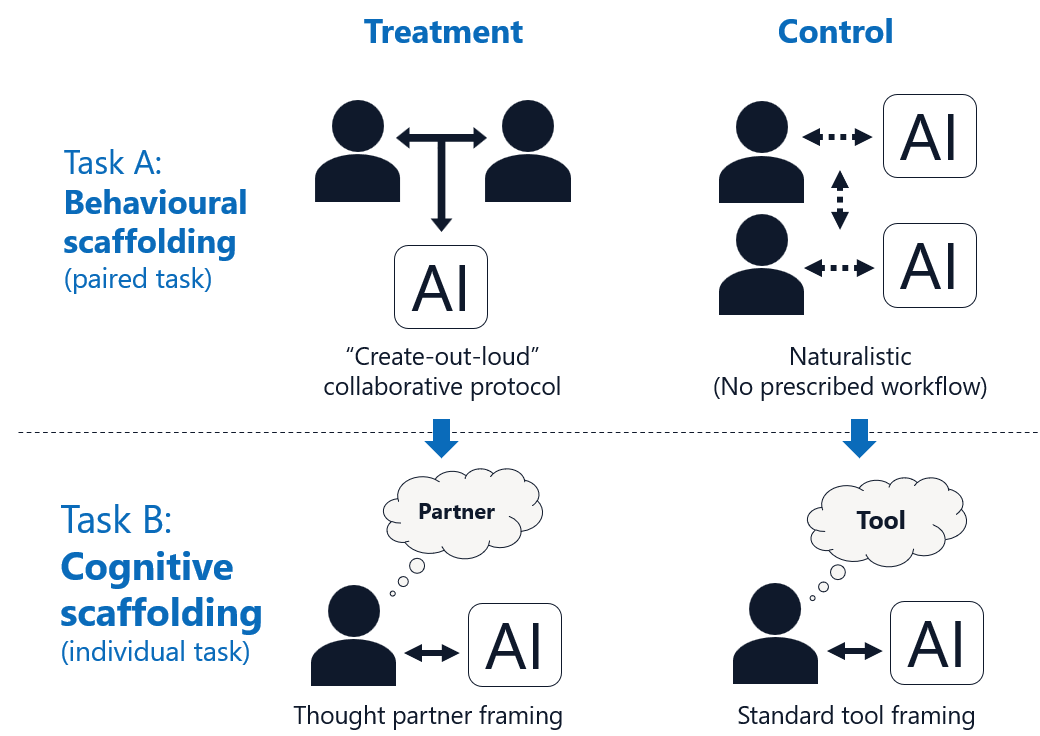}
  \caption{Experimental design. Treatment participants completed Task A using the Create-Out-Loud collaborative protocol and Task B with partnership (thought partner) training. Control participants completed Task A with naturalistic AI use and Task B with standard tool training. Both conditions had identical access to Microsoft Copilot.}
  \label{fig:design}
\end{figure}

Pairs were randomized into two bundled arms: the control arm completed Task A with naturalistic AI use and Task B with basic Copilot training, while the treatment arm completed Task A with the structured protocol and Task B with partnership training. Both arms had full access to Microsoft Copilot. The study tests how people use AI, not whether they have it. Outcomes include LLM-graded document quality, self-reported experience, and belief change measured across the session. We assess all primary outcomes using OLS regression with robust standard errors.

\section{Background and Related Work}
\label{sec:background}

\subsection{Generative AI in Organizations}

The deployment of large language models in enterprise contexts represents a notable change in organizational technology. Unlike previous waves of enterprise software requiring substantial infrastructure and training, GenAI tools can be deployed rapidly with minimal onboarding. Surveys indicated that over 70\% of knowledge workers had access to GenAI tools, yet only 20--30\% reported regular productive use \citep{microsoft2024}.

The heterogeneity in adoption patterns indicates that individual and organizational factors beyond access determine whether AI tools translate to productivity gains. \citet{brynjolfsson2023} find that AI assistance is most beneficial for less-experienced workers, while \citet{dellacqua2023} show that performance depends critically on task--AI alignment. These findings point to the need for structured interventions that help workers navigate the complexities of AI integration.

\subsection{Scaffolding Theory}

The concept of scaffolding originates in developmental psychology \citep{vygotsky1978,wood1976} and has been extended to organizational technology adoption. \citet{edmondson2001} demonstrate that surgical teams adopt new technology more effectively when psychological safety scaffolds enable learning from errors. \citet{leonardi2011} shows that technology affordances are realized only when organizational routines scaffold appropriate usage patterns. Both studies highlight that adoption requires alignment between organizational context and technology demands.

We extend this tradition by distinguishing behavioral scaffolds (structural constraints on interaction patterns) from cognitive scaffolds (interventions targeting mental models). This distinction allows us to test whether different scaffold types have differential effects depending on task structure.

\subsection{A Framework for Scaffold Effectiveness}
\label{sec:framework}

The distinction between behavioral and cognitive scaffolding generates differential predictions depending on task and organizational conditions.

\textbf{Behavioral scaffolding} imposes coordination costs ($C$)---synchronization overhead, communication latency, protocol compliance effort---while potentially generating knowledge aggregation benefits ($B$) through structured information sharing. The net effect depends on $B - C$. \citet{agrawal2024} formalize a related tension: AI increases decision variation, which creates coordination challenges when decisions interact across an organization; coordination mechanisms can facilitate adoption but impose their own costs. We expect behavioral scaffolds to improve outcomes when: (a) partners possess complementary information that structured exchange can surface, (b) the coordination infrastructure is reliable, and (c) the task requires integration across perspectives rather than individual generation. Conversely, we expect behavioral scaffolds to harm outcomes when coordination costs dominate---particularly when infrastructure is unreliable, the task rewards domain-specific depth over breadth, or the AI mediator lacks access to the organizational context needed to synthesize inputs effectively.

\textbf{Cognitive scaffolding} shifts users' mental models from transactional (query-response, one-shot) to dialogic (iterative, conversational). The net effect depends on whether dialogic interaction is more productive for the task. We expect cognitive scaffolds to improve outcomes when: (a) the task benefits from iterative refinement, (b) users' default interaction pattern under-utilizes AI capabilities, and (c) the shift toward dialogue does not impose excessive cognitive load. Cognitive scaffolds require no coordination infrastructure, making them less affected by the organizational frictions that can undermine behavioral approaches.

These considerations suggest that the two scaffold types should have different effect profiles depending on task conditions. Behavioral scaffolds are sensitive to coordination infrastructure and compliance. Cognitive scaffolds operate through individual mental-model shifts and are less dependent on organizational context. We return to this framework when interpreting our results in Section~\ref{sec:cross-task}.

\subsection{Mental Models and Technology Appropriation}

How users conceptualize a technology shapes how they interact with it. \citet{orlikowski1994} introduce ``technological frames'' to describe assumptions and expectations users bring to technology interaction. For generative AI, users may view it as a deterministic tool (expecting reliable, consistent output under their control) or as a probabilistic collaborator (expecting variability and treating outputs as starting points for refinement). This distinction matters because it predicts how users respond to GenAI's inherent variability. Whether users resist or embrace algorithmic judgment depends on framing and experience: people reject algorithms after observing errors \citep{dietvorst2015} but prefer algorithmic advice over human advice when no errors are salient \citep{logg2019}. Our cognitive scaffolding intervention explicitly targets this frame by coaching users to treat AI as an intellectual partner requiring context, iteration, and feedback.

\section{Method}
\label{sec:method}

\subsection{Participants}

A total of 388 employees from Gap Inc.\ participated, organized into 194 pairs (97 per condition). Participants registered for a scheduled ``AI Training Day'' event; research participation was part of the event protocol, and all participants provided informed consent. Selection criteria required full-time employees with access to Microsoft Copilot licenses. Contractors, temporary employees, and research assistants were excluded.

Participants completed a baseline survey capturing demographics, organizational role, tenure, prior AI experience, and participation mode (in-person, remote, or hybrid). Pairs were formed using constrained matching on functional area (coarsened into six categories; see below), baseline AI comfort (Low/Medium/High on a 5-point scale), job level (Individual Contributor, Manager/Senior Manager, Director+), and participation mode (remote or hybrid). When an exact within-category match for functional area was not possible, cross-function pairs were permitted; in such cases, the pair selected one functional area by mutual agreement for the Task A prompt.

\begin{table}[htbp]
\centering
\caption{Participant Characteristics ($N = 388$)}
\label{tab:participants}
\begin{tabular}{lr}
\toprule
Characteristic & $N$ (\%) \\
\midrule
\textbf{Functional area} & \\
\quad Supply Chain, Operations, Strategy \& Governance & 70 (18\%) \\
\quad Stores, Customer Experience, Marketing \& Communications & 85 (22\%) \\
\quad Product, Merchandising, Design \& Development & 58 (15\%) \\
\quad People, Culture \& Workplace & 47 (12\%) \\
\quad Digital, Data \& Technology & 82 (21\%) \\
\quad Finance, Legal \& Risk & 46 (12\%) \\
\midrule
\textbf{Job Level} & \\
\quad Individual Contributor & 163 (42\%) \\
\quad Manager / Senior Manager & 147 (38\%) \\
\quad Director / Senior Director+ & 78 (20\%) \\
\bottomrule
\end{tabular}
\begin{tablenotes}\small
\item \textit{Note:} Functional areas were coarsened into six categories to ensure partners shared functional context during pairing.
\end{tablenotes}
\end{table}

\subsection{Design}

The experiment used a pair-level randomized design. Pairs were randomized into treatment and control using stratified randomization within each functional area stratum, with constrained balance on AI comfort and job level. The AM session served as the Control condition and the PM session as the Treatment condition---a design constraint discussed as a limitation below (see Section~\ref{sec:limitations}).

Both sessions included two sequential tasks designed to be as relevant as possible for participants' work:

\textbf{Task A (Pair Task):} Pairs produced a one-page ``AI Adoption Action Plan'' tailored to their organizational function within a 30-minute time cap. To prevent generic outputs, the task imposed an ``anti-generic'' constraint: every item required either a specific noun (e.g., a named system, dataset, or stakeholder) or a quantitative metric. Each pair produced a single shared document.

\textbf{Task B (Individual Task):} Each participant independently developed a strategic communications response addressing three stakeholder concerns about AI adoption at Gap Inc.: (a) data transparency and privacy, (b) labor displacement, and (c) environmental considerations related to AI energy footprint.

\subsection{Conditions}

\textbf{Control Condition (Naturalistic AI Use + Basic Training):} Pairs completed Task A using their standard working methods. They were permitted to collaborate and use Microsoft Copilot individually, but no specific workflow was enforced. For Task B, individuals received standard functional training on Copilot's interface, capabilities, and basic prompt syntax, treating AI as a software tool requiring correct operation.

\textbf{Treatment Condition (Structured Protocol + Partnership Training):} For Task A, pairs followed a structured ``Create-Out-Loud'' protocol with three sequential steps: (1) meet synchronously via Microsoft Teams, (2) verbally discuss the strategic plan, generating a conversation transcript, and (3) explicitly prompt Microsoft Copilot to draft the initial document based on the transcript (``Copilot-as-Drafter''). For Task B, individuals received partnership training (adapted from the ``AI Mindset'' curriculum developed by Conor Grennan; see Appendix~\ref{app:training} for curriculum details) designed to challenge the ``search engine'' mental model, introduce the ``thought partner'' mental model, use the ``smart intern'' metaphor, and include guided practice in iterative prompting.

\subsection{Measures}

\textbf{Document Quality (Primary).} Documents were graded using GPT-4o-mini (OpenAI, temperature $= 0$) with structured JSON output, following the LLM-as-judge paradigm \citep{zheng2023}. Each document was graded three times independently; the median score was used. Task A used a four-dimension rubric: Opportunities Identified (0--6), Risks Identified (0--6), Action Plan Quality (1--5), and Strategic Insight (1--5), for a maximum of 22 points. Task B used: Problem Understanding (1--5), Internal Strategy (1--5), External Strategy (1--5), and Completeness (1--5), for a maximum of 20 points. Full rubrics with behaviorally anchored examples are in Appendix~\ref{app:rubrics}.

Cross-model validation on a random sample of 176 documents graded by both GPT-4o-mini and GPT-4o yielded: Pearson $r = 0.92$, ICC(2,1) $= 0.92$, mean absolute difference $= 1.81$ points (after correcting cases where the model's reported total did not match the sum of its individual dimension scores; all analyses use dimension sums as the authoritative total). The systematic bias was 0.34 points (GPT-4o-mini slightly higher), consistent across the score range.

\textbf{Human Validation.} A stratified sample of 96 documents (48 Task A, 48 Task B) was independently graded by 8 trained human raters organized in 4 pairs, with each pair grading the same 24 documents. Raters were blinded to condition assignment and LLM scores. Inter-rater reliability and LLM--human agreement results are reported in Section~\ref{sec:grading-integrity} and Appendix~\ref{app:validation}.

\textbf{Self-Reported Experience.} Post-Task A surveys assessed perceived productivity and flow (4 items: productive/efficient, smooth collaboration, confident in output quality, motivated to complete; Cronbach's $\alpha = 0.84$, a measure of internal consistency reliability), perceived Copilot helpfulness (6 items; $\alpha = 0.87$), and future AI intent (3 items; $\alpha = 0.79$). Post-Task B surveys used adapted versions of the same scales: task experience (5 items; the two team-specific items were replaced with individual-task items), Copilot helpfulness (5 items; the collaboration item was dropped since Task B was individual), and future AI intent (3 items, identical). All items used 5-point Likert scales. Full item text is in Appendix~\ref{app:surveys}.

\textbf{Belief Inventories.} Identical inventories at two time points (post-Task A/pre-Task B and post-Task B) measured three constructs: AI as Thought Partner (3 items, ordinal $\alpha = 0.86$--$0.90$), Exploration \& Experimentation (3 items, ordinal $\alpha = 0.93$), and Productivity \& Process (3 items, ordinal $\alpha = 0.89$--$0.90$). All items used 5-point Likert scales. Because no belief inventory was administered before Task A, the first measurement reflects beliefs after participants had already experienced their assigned Task A condition.

\subsection{Analytic Strategy}
\label{sec:analytic}

All primary analyses use intent-to-treat (ITT) comparisons preserving the original randomization. Task A document quality is analyzed at the pair level using OLS with HC2 robust standard errors (since each pair produced one document, pair-level clustering is equivalent to heteroskedasticity-robust inference). Task B document quality is analyzed at the individual level using OLS with CR2 clustered standard errors at the pair level \citep{pustejovsky2018}, accounting for within-pair correlation (ICC $= 0.14$, design effect $= 1.05$, effective $N = 200$ from 210 observations in 155 clusters). Survey outcomes are individual-level responses (each participant completed their own survey, including for the pair task) and use the same CR2 approach.

Belief change is assessed using two specifications. The change-score specification regresses the within-person change from post-Task A to post-Task B beliefs on treatment assignment with covariates and CR2 clustered standard errors at the pair level. The ANCOVA specification regresses post-Task B beliefs on treatment assignment, post-Task A beliefs, and covariates. Both specifications use the same two belief measurements; they differ in functional form. Because the first measurement was collected after Task A (not before randomization), both baselines are post-treatment, and the change-score model captures belief movement during and after Task B, including any recovery from Task A carry-over effects. Multiple comparisons are addressed using Benjamini--Hochberg correction within three outcome families: Survey ITT (3 outcomes), Grading ITT (2 outcomes), and Belief Change (4 outcomes: three subscales plus an overall composite).

Compliance group analyses are reported as exploratory and descriptive. During the experiment, many treatment pairs faced logistical and technical barriers to executing the Create-Out-Loud protocol: some could not connect synchronously, others met but defaulted to individual AI use. To characterize this variation, we constructed a post-hoc compliance classification based on self-reported meeting behavior and Copilot usage patterns. Treatment participants were classified as: \emph{Stranded} (could not meet synchronously; $N = 77$), \emph{Parallel Play} (met but did not use Copilot jointly; $N = 51$), or \emph{True Joint} (met and used Copilot jointly as intended; $N = 37$).\footnote{These counts sum to 165 rather than 194 because compliance classification was based on post-task survey responses. Twenty-nine treatment participants (14.9\%) did not complete the compliance items, so their compliance status could not be determined.}

These classifications reflect post-randomization selection; we report estimated marginal means with 95\% confidence intervals but do not report $p$-values for compliance group comparisons.

\textbf{Attrition Sensitivity.} Because document production was differential by condition (see Section~\ref{sec:balance}), we compute Lee (2009) trimming bounds to characterize the identified set for the ITT estimand under worst-case selection \citep{lee2009}. Under a monotonicity assumption, this procedure trims the observed outcome distribution in the higher-response group from its tails, yielding upper and lower bounds on the treatment effect that are consistent with any pattern of selective attrition. We also model non-production itself as a treatment outcome---logistic regression of document production on treatment assignment---treating attrition as an effect on the extensive margin rather than a nuisance condition.

\textbf{Word-Count Sensitivity.} We report the \citet{oster2019} coefficient stability bound ($\delta$) and a causal mediation model \citep{imai2010} decomposing the total treatment effect into the indirect path through word count (ACME) and the direct path not through word count (ADE), with bootstrap confidence intervals.

\textbf{Session-Effect Sensitivity.} Because treatment and session timing are confounded (AM = Control, PM = Treatment), we report a calibration exercise comparing the observed treatment effect size against the range of circadian performance effects documented in the cognitive psychology literature \citep{monk2005,blatter2007,valdez2012}. We compute the residual treatment effect under a range of assumed session-effect magnitudes and report the breakeven point.

This study was not pre-registered. Covariates were pre-specified based on theoretical relevance \citep{lin2013}. Appendix~\ref{app:transparency} provides a transparency table distinguishing pre-specified from post-hoc analytical decisions.

Post-task surveys were completed by 370 of 388 participants (95.4\%). Of these, 270 (73\%) were linked to pair IDs. All survey and belief-change analyses use this linked subset ($N = 270$), because CR2 clustering requires pair membership. The 27\% unlinked were balanced on treatment status (Fisher exact $p = .638$, logistic regression OR $= 0.875$, $p = .571$) and covariates (all $p > .22$).

\section{Results}
\label{sec:results}

\subsection{Balance and Manipulation Checks}
\label{sec:balance}

Randomization balanced baseline characteristics across conditions. No significant differences emerged in functional area distribution, job level, AI experience categories, baseline AI comfort, or organizational tenure (all $p > .40$).

\textbf{Differential Attrition.} Although baseline characteristics were balanced, document production was not. Attrition from the randomized sample to the analysis sample differed significantly by condition:

\begin{table}[htbp]
\centering
\caption{Differential Attrition by Task and Condition}
\label{tab:attrition}
\begin{threeparttable}
\footnotesize
\begin{tabular}{lcccccc}
\toprule
Task & Ctrl: Rand $\to$ Prod & Trt: Rand $\to$ Prod & Ctrl \% & Trt \% & Fisher $p$ & OR \\
\midrule
Task A (pairs) & 97 $\to$ 93 & 97 $\to$ 71 & 4.1 & 26.8 & $< .001$ & 8.43 \\
Task B (indiv.) & 194 $\to$ 123 & 194 $\to$ 87 & 36.6 & 55.2 & $< .001$ & 2.13 \\
\bottomrule
\end{tabular}
\end{threeparttable}
\end{table}

Non-production was high even among controls (37\% for Task B) reflecting a mix of participant attrition from the event and non-submission among those who stayed. Task B was the final exercise of a full-day training event with some participants leaving before or during the task, while others remained but did not submit a document within the allotted time. The \emph{differential} attrition between conditions is the identification concern. For Task A, treatment pairs were over 8 times more likely to fail to produce a document, likely because the coordination requirements of the Create-Out-Loud protocol made it harder to complete work within the allotted time. For Task B, attrition was significantly higher in the treatment condition (55\% vs.\ 37\%), though the mechanism is less clear. Treatment participants had just completed the more demanding Task A protocol, and fatigue or frustration from that experience may have carried over. We address this differential attrition formally using Lee (2009) trimming bounds in Section~\ref{sec:taskA}.

\subsection{Task A: Document Quality (Pair-Level)}
\label{sec:taskA}

In Task A, pairs collaboratively produced an AI Adoption Action Plan for their organizational function. Treatment pairs followed the Create-Out-Loud protocol (synchronous discussion, then joint AI drafting); control pairs used Copilot however they chose.

\textbf{ITT Analysis.} Treatment pairs, instructed to work jointly with AI, scored significantly lower than control pairs, which worked without any enforced workflow:

\begin{table}[htbp]
\centering
\caption{Task A Document Quality by Condition}
\label{tab:taskA-itt}
\begin{tabular}{lccc}
\toprule
 & $N$ (pairs) & Mean & SD \\
\midrule
Control (AM) & 93 & 15.63 & 5.92 \\
Treatment (PM) & 71 & 10.68 & 6.34 \\
\textbf{Difference} & & \textbf{$-$4.96} & \\
\bottomrule
\end{tabular}
\end{table}

OLS with HC2 robust standard errors: $b = -4.958$, SE $= 0.888$, 95\% CI $[-6.71, -3.20]$, $p < .001$, $|d| = 0.81$. The effect survives Benjamini--Hochberg correction (BH-adjusted $p = 1.97 \times 10^{-7}$).

The estimated treatment difference was uniform across all four rubric dimensions:

\begin{table}[htbp]
\centering
\caption{Task A Treatment Effects by Rubric Dimension}
\label{tab:taskA-dimensions}
\begin{tabular}{lccc}
\toprule
Dimension (scale) & Control Mean & Treatment Mean & Difference \\
\midrule
Opportunities (0--6) & 4.57 & 2.99 & $-$1.58 \\
Risks (0--6) & 4.06 & 2.48 & $-$1.58 \\
Action Plan (1--5) & 3.59 & 2.69 & $-$0.90 \\
Insight (1--5) & 3.41 & 2.52 & $-$0.89 \\
\bottomrule
\end{tabular}
\end{table}

Control pairs also produced more document versions (3.28 vs.\ 2.44 mean versions) and longer documents (740 vs.\ 454 mean words).

\textbf{Lee Bounds.} Given severe differential attrition (4.1\% control vs.\ 26.8\% treatment), we computed \citet{lee2009} trimming bounds by trimming 23.7\% of control observations from the tails of the control outcome distribution:

\begin{table}[htbp]
\centering
\caption{Lee (2009) Trimming Bounds: Task A}
\label{tab:lee-taskA}
\begin{tabular}{lcc}
\toprule
Bound & Estimate & Bootstrap 95\% CI \\
\midrule
Lower (most conservative) & $-$7.01 & [$-$8.70, $-$5.46] \\
Point estimate (naive ITT) & $-$4.96 & --- \\
Upper (least conservative) & $-$3.34 & [$-$5.24, $-$1.62] \\
\bottomrule
\end{tabular}
\end{table}

Both bounds are negative and statistically significant. The sign and significance of the treatment effect are robust to worst-case selective attrition.

\textbf{Non-production as outcome.} Modeling document production itself as a treatment outcome on the full randomized sample: treatment pairs were significantly less likely to produce a document (odds ratio [OR] $= 0.12$, 95\% CI $[0.04, 0.35]$, $p < .001$). An OR below 1 indicates lower odds of the outcome in the treatment group; here, treatment pairs had roughly one-eighth the odds of producing a document compared to control pairs. The structured protocol appears to have created barriers to task completion, an effect on the extensive margin that accompanies the intensive-margin quality difference.

\textbf{Word-Count Sensitivity.} The LLM grader exhibits strong word-count bias: Spearman $\rho = 0.646$ between total score and document word count for Task A. When word count is added as a covariate, the treatment effect reduces from $-4.96$ to $-3.34$, a 33\% reduction. The \citet{oster2019} coefficient stability bound: $\delta = 3.67$ under the recommended $R^2_{\max} = 1.3 \times R^2_{\text{controlled}}$, and $\delta = 0.65$ under $R^2_{\max} = 1$. The $\delta = 3.67$ under the standard assumption indicates that unobservable confounding would need to be nearly four times as important as the observed word-count confound to explain the entire treatment effect. However, $\delta = 0.65$ under the most conservative assumption ($R^2_{\max} = 1$) falls below 1, meaning that if the true $R^2$ of a fully specified model were close to 1, unobservables proportional to word count could plausibly account for the effect. A formal causal mediation analysis \citep{imai2010} estimated that 32.6\% of the total effect operates through the word-count channel (ACME $= -1.61$, 95\% CI $[-2.59, -0.80]$, $p < .001$), while the remaining direct effect is $-3.34$ (ADE, 95\% CI $[-4.95, -1.72]$) and remains statistically significant.

\textbf{Session-Effect Sensitivity.} Because treatment and control conditions were assigned to different sessions (AM = Control, PM = Treatment), the treatment effect estimate is confounded with time-of-day. If afternoon participants were more fatigued or less motivated, some or all of the observed difference could reflect session timing rather than the intervention itself. We investigate how large such a session effect would need to be to fully account for our findings. The observed Task A effect ($|d| = 0.81$) would require a pure session effect of $d = 0.81$ to be fully explained by PM fatigue---approximately 3 times larger than the largest circadian performance decrements documented in the cognitive psychology literature ($d \approx 0.20$--$0.30$; \citealp{blatter2007,valdez2012}). Even under the most conservative literature-calibrated assumption ($d = 0.30$), the residual treatment effect remains statistically significant. Within the PM session, compliance subgroups show a 2.7-point spread in mean scores, suggesting protocol-related variation beyond time-of-day. The Task B result provides a partial falsification: treatment participants (also PM) outperformed control (AM) on the binary outcome (OR $= 2.07$), contradicting a uniform session-effect explanation (see Appendix~\ref{app:session} for the full sensitivity table).

\textbf{Compliance Patterns (Descriptive).} As described in Section~\ref{sec:analytic}, many treatment pairs could not execute the Create-Out-Loud protocol as intended. Based on self-reported meeting behavior and Copilot usage, treatment pairs were classified post-hoc as \emph{True Joint} (met synchronously and used Copilot jointly), \emph{Parallel Play} (met but used Copilot individually), or \emph{Stranded} (could not meet synchronously). Among treatment pairs with available compliance data, estimated marginal means by compliance group were:

\begin{table}[htbp]
\centering
\caption{Task A Compliance Group Means (Descriptive)}
\label{tab:compliance}
\begin{threeparttable}
\begin{tabular}{lcc}
\toprule
Group & $n$ & Mean \\
\midrule
Control & 84 & 17.2 \\
True Joint & 22 & 12.5 \\
Stranded & 23 & 12.3 \\
Parallel Play & 17 & 9.8 \\
\bottomrule
\end{tabular}
\begin{tablenotes}\small
\item \textit{Note:} Descriptive only; compliance group membership reflects post-randomization selection. Compliance data were available for 146 of 164 pairs with documents.
\end{tablenotes}
\end{threeparttable}
\end{table}

All three treatment compliance subgroups scored below control, with a 2.7-point spread within the treatment group (Parallel Play: 9.8 vs.\ True Joint: 12.5). This within-treatment variation suggests that protocol execution, not just time of day, contributed to the outcome pattern.

\textbf{Time-to-quality.} We examined version-level quality trajectories using document snapshots captured at $T = 10$, $T = 20$, and $T = 30$ minutes. Treatment pairs were half as likely to reach the 70\% quality threshold as controls (33.8\% vs.\ 68.8\%; Fisher exact $p < .0001$), though among those who reached it, the mean version was similar (3.08 vs.\ 3.12). The structured protocol slowed not just final quality but also the rate at which quality accumulated.

\subsection{Task B: Document Quality (Individual-Level)}
\label{sec:taskB}

In Task B, each participant independently developed a strategic communications response addressing three stakeholder concerns about AI adoption. Treatment individuals received partnership training reframing AI as a thought partner; control individuals received standard Copilot training. The pre-specified primary outcome was continuous document quality.

\textbf{Continuous ITT Analysis.} Task B exhibited severe ceiling effects: 68.1\% of all documents scored 20/20 (the maximum). The task was more constrained than Task A (participants addressed three specific stakeholder concerns rather than generating an open-ended strategy document) and with Copilot available, comprehensive responses could be produced with relatively little iteration. The LLM grader's leniency at the top of the scale likely compounded the saturation.

\begin{table}[htbp]
\centering
\caption{Task B Continuous Document Quality by Condition}
\label{tab:taskB-itt}
\begin{tabular}{lccc}
\toprule
 & $N$ & Mean & SD \\
\midrule
Control (AM) & 123 & 17.3 & 3.44 \\
Treatment (PM) & 87 & 18.1 & 3.60 \\
\bottomrule
\end{tabular}
\end{table}

OLS with CR2 clustered standard errors: $p = .223$. BH-adjusted $p = .223$. Treatment participants scored slightly higher on average (18.1 vs.\ 17.3), but the difference was not statistically significant. With the majority of scores concentrated at the ceiling, the continuous model lacked the variance to detect treatment effects even if the intervention shifted the probability of reaching the top of the distribution.

\textbf{Lee Bounds (Continuous).} Because attrition was differential (36.6\% control vs.\ 55.2\% treatment), the continuous null could reflect selective survival rather than a true absence of treatment effects. Lee bounds on the continuous ITT range from $-1.58$ to $+1.90$, spanning zero, consistent with the null but not ruling out effects in either direction under worst-case attrition.

\textbf{Binary Model.} When 68\% of observations are at the ceiling, a continuous model cannot detect differences in the probability of reaching the top. We estimated a binary logistic model: did the participant produce a perfect-score document (20/20)?

\begin{table}[htbp]
\centering
\caption{Task B Binary Logistic Model: Threshold Sensitivity}
\label{tab:taskB-binary}
\begin{tabular}{lccccc}
\toprule
Threshold & Control & Treatment & OR & CR2 95\% CI & CR2 $p$ \\
\midrule
Score $= 20$ & 61.8\% (76/123) & 77.0\% (67/87) & \textbf{2.07} & [1.12, 3.83] & \textbf{.022} \\
Score $\geq 19$ & --- & --- & \textbf{2.00} & [1.08, 3.70] & \textbf{.029} \\
Score $\geq 18$ & 64.2\% (79/123) & 77.0\% (67/87) & \textbf{1.87} & [1.01, 3.45] & \textbf{.049} \\
Score $\geq 17$ & --- & --- & 1.61 & [0.87, 2.99] & .131 \\
\bottomrule
\end{tabular}
\end{table}

The OR of 2.07 corresponds to a 15.2 percentage-point increase in the probability of achieving a perfect score (from 61.8\% to 77.0\%), or a risk ratio of 1.25.

\textbf{Lee Bounds (Binary $\geq 20$).} Given the positive binary result, we tested whether it survives worst-case selective attrition. Lee bounds on the binary outcome range from $-0.10$ to $+0.31$. The lower bound crosses zero, meaning the binary effect cannot be confirmed as robust to worst-case attrition. This is an important caveat on the OR $= 2.07$ finding.

\emph{Note on multiple comparisons:} The binary model is not included in the Grading ITT BH family. If added to a 3-test family, the BH-adjusted $p$ would be .033, still below .05.

\textbf{Time-to-quality.} Task B showed no significant differences in time-to-quality: 89.7\% of treatment participants and 86.2\% of controls reached the 70\% threshold, with nearly identical mean versions (2.36 vs.\ 2.39).

\subsection{Survey Outcomes}
\label{sec:survey}

Survey outcomes fall into two categories: post-task self-reports collected after Task A (experience, Copilot helpfulness, and future AI use intent) and belief change measured between the two post-task belief inventories (post-Task A and post-Task B). Task B collected adapted versions of the same experience and helpfulness scales, but the primary survey outcome for the cognitive scaffolding intervention is belief change, reported below.

\textbf{Self-Reported Experience (Post-Task A ITT).} No significant differences emerged between treatment and control on Task A post-task survey outcomes:

\begin{table}[htbp]
\centering
\caption{Post-Task A Survey ITT Results}
\label{tab:survey-itt}
\begin{threeparttable}
\begin{tabular}{lccccc}
\toprule
Scale & Ctrl Mean & Trt Mean & $d$ & Raw $p$ & BH $p$ \\
\midrule
Experience & 3.38 & 3.50 & 0.14 & .505 & .757 \\
Copilot Helpfulness & 3.84 & 3.77 & 0.08 & .254 & .757 \\
Future Use Likelihood & 4.16 & 4.02 & 0.16 & .797 & .797 \\
\bottomrule
\end{tabular}
\begin{tablenotes}\small
\item \textit{Note:} $p$-values from covariate-adjusted OLS with CR2 clustered standard errors; BH correction applied within the Survey ITT family.
\end{tablenotes}
\end{threeparttable}
\end{table}

\textbf{Belief Change (Pre-Post).} We assessed belief change using a difference-in-differences approach. The overall belief composite showed a significant DiD interaction ($p = .023$, BH-adjusted $p = .047$).

As a robustness check, we estimated a cumulative link mixed model (CLMM; ordinal logistic regression). The time $\times$ group DiD interaction was significant: $\beta = 0.582$, SE $= 0.153$, $z = 3.81$, $p < .001$, cumulative OR $= 1.79$, 95\% CI $[1.33, 2.41]$.

\begin{table}[htbp]
\centering
\caption{Belief Change by Dimension (DiD Specification)}
\label{tab:beliefs}
\begin{threeparttable}
\begin{tabular}{lccccc}
\toprule
Dimension & Trt $\Delta$ & Ctrl $\Delta$ & Raw $p$ & BH $p$ & $d$ \\
\midrule
Exploration \& Experimentation & $+$0.26 & $-$0.02 & .003 & \textbf{.013} & 0.41 \\
Overall (composite) & --- & --- & .023 & \textbf{.047} & --- \\
AI as Thought Partner & $+$0.24 & $+$0.08 & .147 & .150 & 0.34 \\
Productivity \& Process & $+$0.17 & $+$0.10 & .150 & .150 & 0.23 \\
\bottomrule
\end{tabular}
\begin{tablenotes}\small
\item \textit{Note:} Raw $p$-values from covariate-adjusted OLS on change scores with CR2 clustered standard errors. BH correction applied within the Belief Change family (4 outcomes). Cohen's $d$ values are signed: positive indicates greater change in the treatment group.
\end{tablenotes}
\end{threeparttable}
\end{table}

Only Exploration \& Experimentation survived BH correction at the individual subscale level (BH $p = .013$). The overall belief composite also shifted significantly (BH $p = .047$), driven primarily by the Exploration dimension. Post-hoc power analysis suggests the Productivity comparison was inadequately powered (achieved power $= 0.35$, requiring $N = 307$ per group for 80\% power at the observed effect size).

\begin{figure}[htbp]
  \centering
  \includegraphics[width=\textwidth]{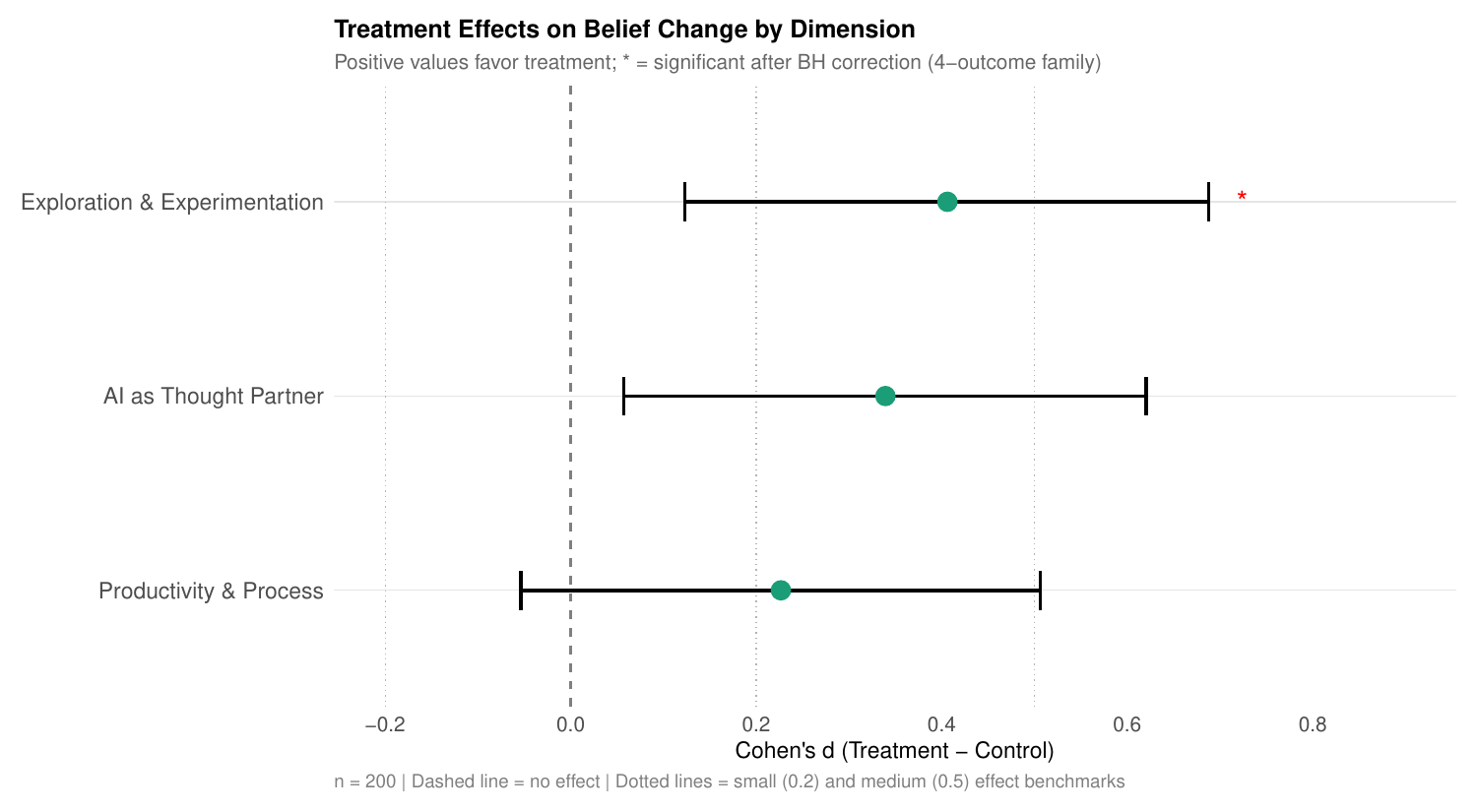}
  \caption{Treatment effects on belief change by dimension (Cohen's $d$). Positive values indicate greater belief change in the treatment group. Asterisks indicate significance after BH correction within the 4-outcome Belief Change family. Only Exploration \& Experimentation survives correction (BH $p = .013$).}
  \label{fig:forest}
\end{figure}

\begin{figure}[htbp]
  \centering
  \includegraphics[width=\textwidth]{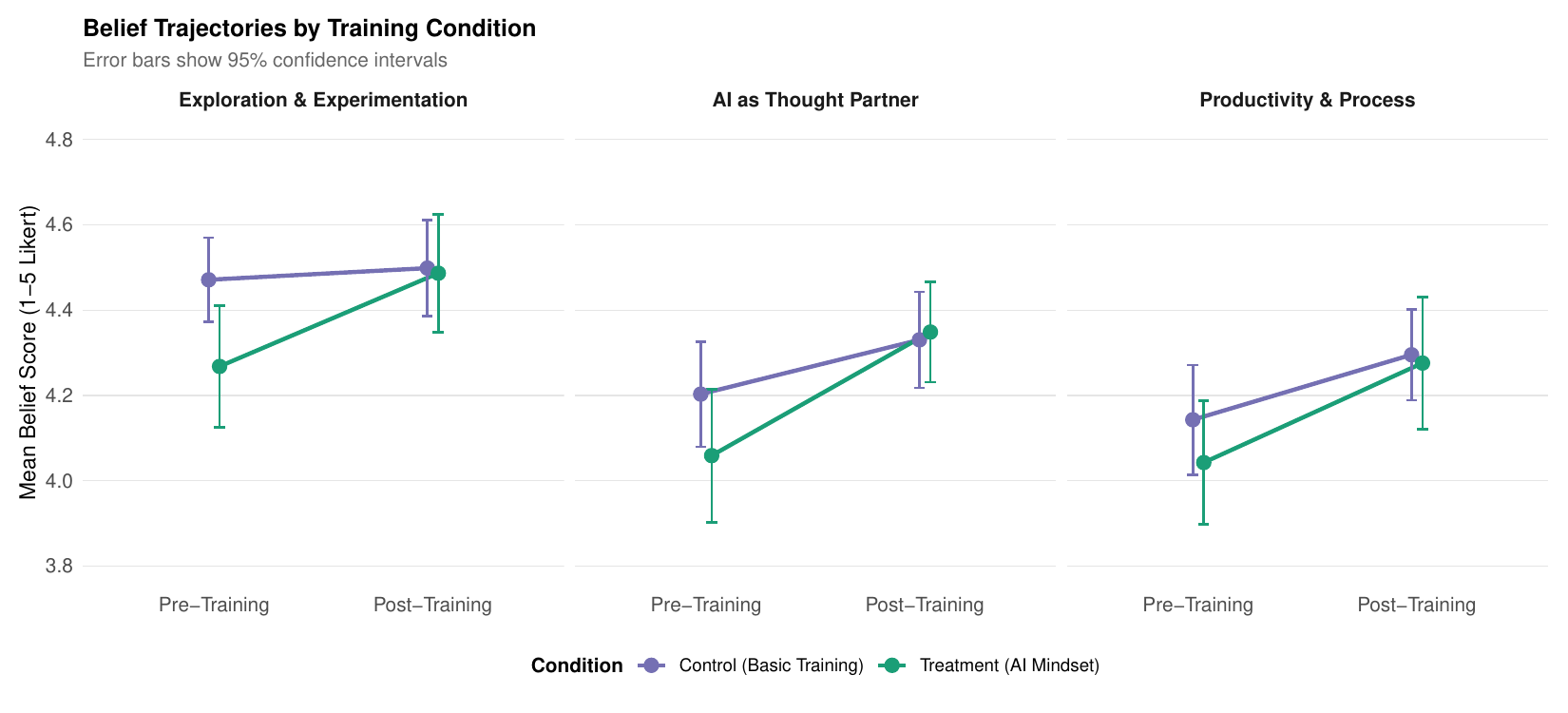}
  \caption{Belief trajectories by training condition. Post-Task A and Post-Task B measurements shown. Error bars show 95\% confidence intervals. Treatment participants start lower due to carry-over from Task A friction but show steeper gains.}
  \label{fig:trajectories}
\end{figure}

\begin{figure}[htbp]
  \centering
  \includegraphics[width=\textwidth]{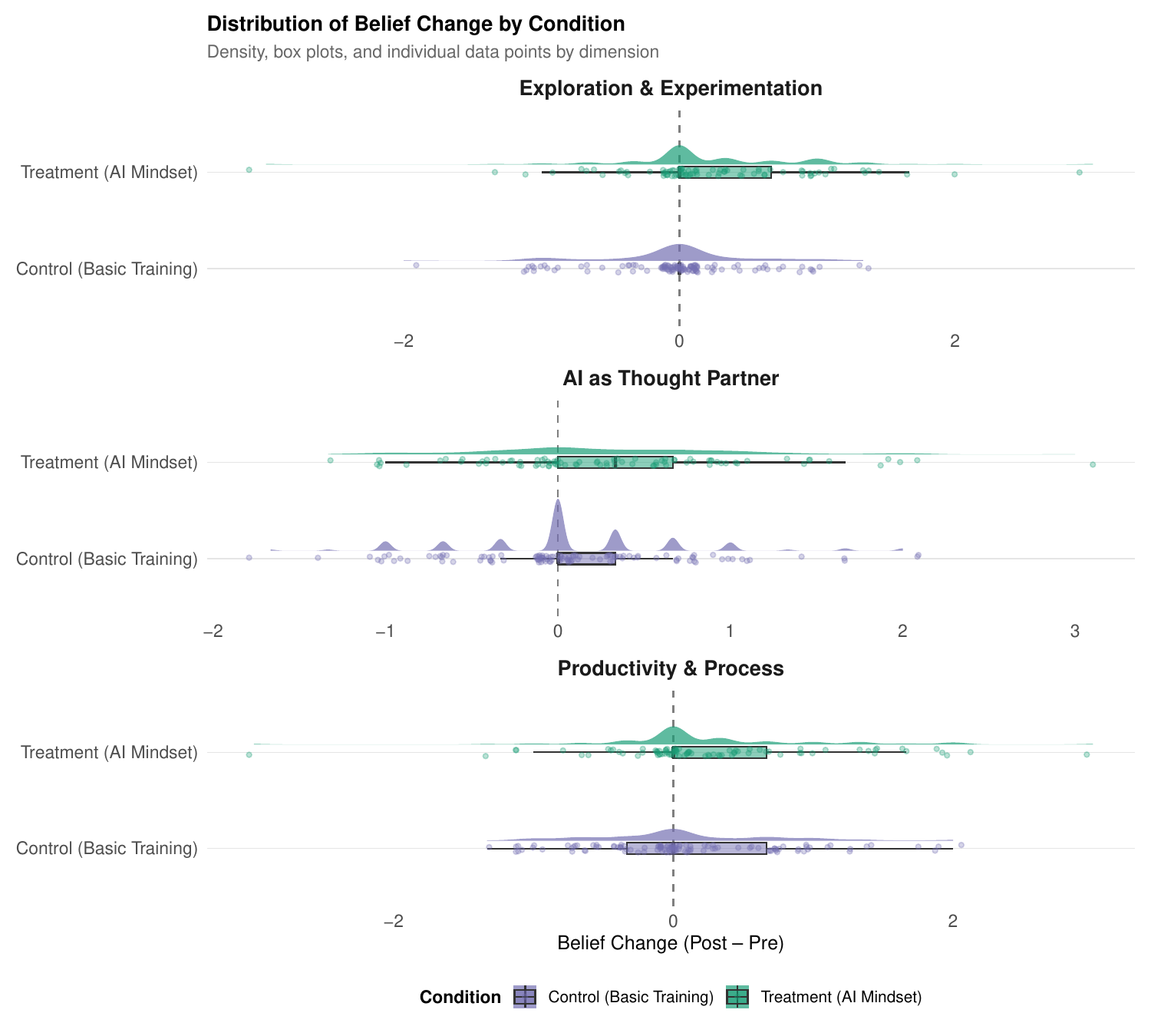}
  \caption{Distribution of belief change (Post-Task B -- Post-Task A) by condition. Raincloud plots show full distribution, box plots, and individual data points.}
  \label{fig:raincloud}
\end{figure}

\textbf{Interpretive caveat and specification hierarchy.} Treatment participants entered Task B with lower belief scores than control participants ($d = 0.34$, $p = .008$), reflecting a carry-over effect from Task A friction. Because beliefs were first measured after Task A (not before randomization), both specifications use a post-treatment baseline. The change-score specification captures belief movement from post-Task A to post-Task B, which includes both any recovery from the Task A depression and any genuine Task B training effect. This is our primary specification.

The \textbf{ANCOVA specification} controls for post-Task A beliefs on treatment assignment and post-Task A beliefs, adjusting for the level at which participants entered Task B. Because the post-Task A baseline is itself post-treatment, conditioning on it risks collider bias \citep{angrist2009}; we report it as a sensitivity check rather than a primary specification. The ANCOVA was not significant for any dimension, which is informative: it suggests that after adjusting for the depressed post-Task A starting point, the observed change-score effects likely reflect recovery toward pre-intervention levels rather than genuine training-induced belief shifts. This is consistent with the training transfer literature, where meta-analytic evidence indicates that brief interventions often produce immediate attitudinal shifts that decay without sustained reinforcement \citep{blume2010}. We report both specifications transparently.

\subsection{LLM Grading Integrity}
\label{sec:grading-integrity}

The LLM grader exhibited a systematic positive bias relative to human raters ($+4.9$ points; ICC $= 0.56$), but rank ordering was preserved for Task A (Spearman $\rho = 0.79$) and moderate for Task B ($\rho = 0.58$). Crucially, the bias was symmetric across conditions (Task A: Wilcoxon $p = .18$; Task B: $p = .12$), meaning treatment effect estimates are not distorted by differential grading error. Human-human agreement was similarly modest (ICC $= 0.47$), suggesting rubric ambiguity rather than LLM-specific failure. Full grading integrity results, including determinism and score corrections, are in Appendix~\ref{app:validation}.

\section{Discussion}
\label{sec:discussion}

\subsection{Summary of Findings}

The experiment produced three principal findings:

\begin{enumerate}
  \item \textbf{Mandating structured collaborative AI protocols was associated with lower document quality among pairs that produced documents} (Task A: $b = -4.96$, $p < .001$, confounded with time of day), \textbf{and with substantially lower document production} (OR $= 0.12$). The Create-Out-Loud protocol was associated with lower-scoring documents than allowing pairs to use AI however they naturally would. This effect was uniform across rubric dimensions and partially attributable to the LLM grader's word-count sensitivity (33\% reduction when controlling for length). The quality analysis conditions on document production and thus estimates the intensive-margin effect; the full organizational cost includes both the quality difference and the production shortfall.

  \item \textbf{Partnership training was associated with higher odds of producing a top-quality individual document} in an exploratory binary model (OR $= 2.07$, $p = .022$, CR2 clustered SEs), though the pre-specified continuous ITT model was null ($p = .223$). The binary threshold was chosen post-hoc in response to observed ceiling saturation (Appendix~\ref{app:transparency}). The effect was consistent across thresholds of 18--20 but attenuated below 18.

  \item \textbf{Partnership training was associated with greater positive change in exploration beliefs} (BH-adjusted $p = .013$) and the overall belief composite (BH-adjusted $p = .047$), though the AI as Thought Partner subscale did not reach significance after correction (BH $p = .150$). The ANCOVA specification was not significant for any dimension, suggesting that observed changes likely reflect recovery from the post-Task A belief depression rather than genuine training-induced shifts.
\end{enumerate}

\subsection{Interpreting the Task A Effect}

The Task A result requires careful interpretation given three considerations.

First, the LLM grader exhibits substantial word-count sensitivity ($\rho = 0.65$), and control pairs wrote 63\% longer documents (740 vs.\ 454 words). The grader may be penalizing conciseness rather than detecting lower quality. Human validation confirmed a systematic positive LLM calibration bias but found that treatment effect direction was preserved under human scoring (Section~\ref{sec:grading-integrity}).

Second, the protocol imposed real coordination costs. It required synchronous meetings, verbal discussion, and sequential AI prompting, effort that control pairs could instead direct toward content. The protocol positioned AI as the central drafter working from a conversation transcript, but AI lacks access to participants' organizational context. This pattern is consistent with \citet{kellogg2020} finding that algorithmic coordination mechanisms can generate worker resistance when they restrict autonomy, and with meta-analytic evidence that human--AI combinations underperform the best of humans or AI alone in most task domains \citep{vaccaro2024}.

Third, many treatment pairs could not execute the protocol at all. Of the 62 treatment pairs with compliance data, 23 (37\%) were classified as Stranded. The ITT estimate blends the effect of the protocol itself with the cost of failed coordination.

The design confound between treatment and time of day (AM = Control, PM = Treatment) must also be acknowledged. However, comparisons within the treatment group provide some leverage: the variation across compliance subgroups suggests the protocol itself contributed to the outcome pattern.

\subsection{Task B: Ceiling Effects and the Binary Model}
\label{sec:taskB-discussion}

The continuous null result on Task B ($p = .223$) is difficult to interpret because the rubric could not discriminate among the majority of participants. When 68\% score 20/20, a linear model lacks variance to detect treatment effects. The binary model estimates a statistically significant association between training condition and document quality (OR $= 2.07$, 95\% CI $[1.12, 3.83]$, $p = .022$). However, this finding should be interpreted cautiously: the continuous model is null, the binary threshold was chosen post-hoc, and the effect is not robust below score $\geq 18$, and differential attrition is an additional concern (see Section~\ref{sec:limitations}).

Human validation confirmed that human raters used a wider range of the scale. The mean human score for Task B was 11.8 compared to the LLM's 17.3, indicating that the LLM's generous scoring amplified the ceiling effect.

\subsection{Two Interventions in Two Contexts}
\label{sec:cross-task}

The Task A and Task B results should be read as two bounded operational lessons rather than as evidence for a general scaffold taxonomy. The two interventions differ not only in scaffold type (behavioral vs.\ cognitive) but also in unit of analysis (pair vs.\ individual), task structure (open-ended vs.\ constrained), outcome saturation (no ceiling vs.\ severe ceiling), attrition rates, and temporal sequencing. Attributing the pattern solely to the behavioral-vs.-cognitive distinction would be overinterpreting a confounded comparison. Task A speaks to the risks of mandating synchronous AI-mediated protocols under infrastructure friction; Task B speaks to the potential of partnership-style training for individual AI appropriation. Each finding is bounded by its specific design features.

With that caveat, the framework from Section~\ref{sec:framework} helps organize the pattern of results:

\textbf{Behavioral scaffolding} was associated with lower output quality, consistent with coordination costs exceeding collaboration benefits. The framework suggests behavioral scaffolding should help when compliance is high and the task requires cross-perspective integration. In our study, compliance was low (37\% of treatment pairs were Stranded), infrastructure was unreliable, and the task rewarded domain-specific depth---conditions under which the framework anticipates net harm from imposed coordination.

\textbf{Cognitive scaffolding} showed a different pattern. The framework suggests cognitive scaffolds should improve outcomes when users' default interaction patterns under-utilize AI capabilities. The continuous ITT was null ($p = .223$), but the exploratory binary model (OR $= 2.07$, $p = .022$) suggests a positive effect attenuated by ceiling saturation, consistent with cognitive reframing helping users reach the top of the quality distribution rather than shifting the entire distribution. Treatment participants showed greater positive belief change (Exploration BH $p = .013$), but the ANCOVA null for all dimensions suggests this reflects recovery from Task A carry-over rather than durable reframing. The belief results are therefore consistent with this logic. Cognitive scaffolding may shift mental models but the evidence for genuine training-induced change is not strong enough to confirm it.

One possible reading---that cognitive reframing may be more tractable than behavioral mandates---should be treated as a hypothesis for future testing. This interpretation aligns with \citeauthor{lebovitz2022}'s (\citeyear{lebovitz2022}) finding that ``engaged augmentation'' with AI emerged naturally in organizational units with the right cognitive orientation but could not be mandated through structural interventions alone.

We examined treatment effect heterogeneity by prior AI experience. The visual pattern of divergence at the ``Experienced'' level does not reach significance with CR2 clustered standard errors (interaction $F = 0.16$, $p = .855$).

\begin{figure}[!t]
  \centering
  \includegraphics[width=\textwidth]{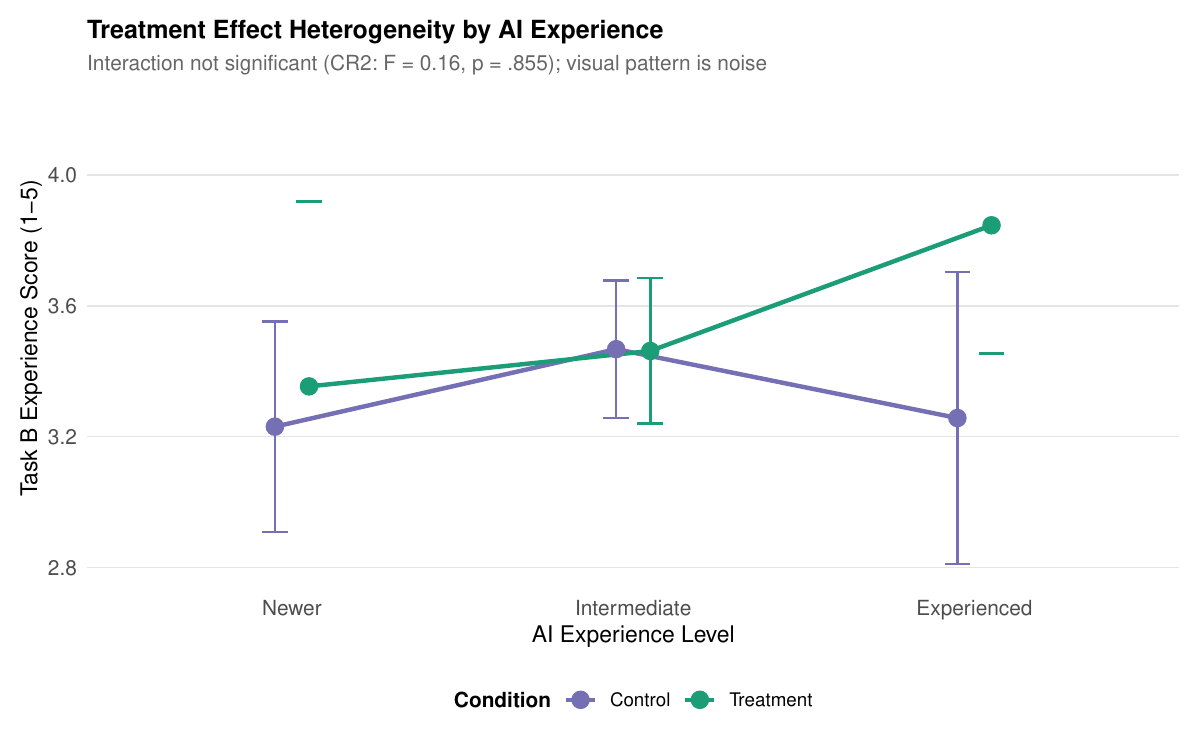}
  \caption{Treatment effect heterogeneity by prior AI experience. The visual pattern of divergence at the ``Experienced'' level does not reach significance with CR2 clustered standard errors (interaction $F = 0.16$, $p = .855$).}
  \label{fig:heterogeneity}
\end{figure}

\subsection{Limitations}
\label{sec:limitations}

\textbf{Differential Attrition.} Treatment pairs were substantially less likely to produce documents for both tasks, meaning the quality estimates condition on a selected sample. Lee trimming bounds (reported in Section~\ref{sec:taskA}) confirm that the Task A quality effect survives worst-case selective attrition, but the Task B binary effect cannot be distinguished from a selection artifact.

\textbf{AM/PM Confound.} Treatment and control conditions were administered at different times of day. The calibration exercise in Appendix~\ref{app:session} compares the observed effect against circadian performance decrements ($d \approx 0.20$--$0.30$; \citealp{blatter2007,valdez2012}), but the session-timing confound extends beyond circadian physiology. Operational differences between AM and PM windows (IT infrastructure load, facilitator fatigue, room logistics, or throughput constraints) could also differ systematically and are not captured by the circadian calibration. We cannot fully rule out session-specific artifacts as a contributor to the Task A result.

\textbf{Carry-Over Effects.} Participants experienced both tasks sequentially, and beliefs were first measured only after Task A---not before randomization. Treatment participants who experienced Task A friction entered Task B with depressed beliefs relative to control ($d = 0.34$, $p = .008$). The change-score results may therefore reflect recovery from this depression rather than a genuine training effect; the null ANCOVA results are consistent with this interpretation.

\textbf{Word-Count Bias in LLM Grading.} The LLM grader's word-count sensitivity ($\rho = 0.65$ for Task A) is a substantial measurement concern. The Oster coefficient stability bound exceeds 1 under the standard assumption ($\delta = 3.67$ at $R^2_{\max} = 1.3 \times R^2_{\text{controlled}}$) but falls below 1 under the most conservative assumption ($\delta = 0.65$ at $R^2_{\max} = 1$), meaning that under conservative conditions, unobservables proportional to word count could plausibly account for the full effect. Human validation was consistent with symmetric LLM calibration bias across conditions.

\textbf{Self-Reported Compliance.} Compliance classification relied on self-reported behavior. This is why compliance analyses are descriptive only.

\textbf{Single Organization and Session.} All participants came from Gap Inc.\ and completed both tasks in a single day.

\textbf{LLM Grading as Novel Methodology.} Human validation (61 of 96 assigned documents with complete dual ratings from four rater pairs) revealed systematic positive calibration bias (mean difference $+4.9$ points; ICC $= 0.56$, Krippendorff's $\alpha = 0.46$). LLM--human rank-order agreement was strong for Task A (Spearman $\rho = 0.79$) but moderate for Task B ($\rho = 0.58$), which may mean that the Task B grading rubric may not discriminate well across the range of document quality---consistent with the ceiling effect that concentrates 68\% of scores at the maximum. This lower rank-order fidelity for Task B reinforces our framing of the binary threshold as an exploratory Task B outcome and the continuous ITT as difficult to interpret. Human-human agreement was also modest (ICC $= 0.47$, weighted $\kappa = 0.21$), indicating that rubric ambiguity contributes substantially to measurement noise beyond LLM-specific limitations.

\textbf{Power.} The study detects effects of $d > 0.40$ at 80\% power. Survey null results may reflect real effects below the detection floor.

\section{Practical Implications}
\label{sec:implications}

\textbf{For AI deployment strategy:} Access to AI tools is necessary but insufficient. Our results suggest that mandating rigid collaborative AI protocols may carry risks. Organizations should consider piloting and evaluating such protocols before broad deployment.

\textbf{For training design:} Partnership training was associated with higher individual output quality at the top of the distribution. Our results are consistent with the idea that cognitive scaffolding for individual AI use may warrant attention before investing in structured team protocols, though this inference rests on a cross-task comparison with multiple confounds.

\textbf{For measurement:} The Task B ceiling effect and Task A word-count sensitivity illustrate that how we measure AI-augmented work quality can substantially affect conclusions.

\section{Conclusion}
\label{sec:conclusion}

This experiment tested two approaches to helping employees use AI more effectively: mandating a structured collaborative protocol, and reframing AI through partnership training. The results complicate simple narratives about AI adoption.

Pairs assigned to the structured collaborative protocol (PM session) produced lower-scoring documents than those working naturally (AM session), a finding complicated by the AM/PM confound, the LLM grader's word-count sensitivity, and high non-compliance. The partnership training condition was associated with higher odds of producing top-quality individual work (in the exploratory binary model, OR $= 2.07$, $p = .022$). Treatment participants also showed greater positive belief change, supported by both OLS DiD and ordinal CLMM (OR $= 1.79$, $p < .001$), but the ANCOVA null and the carry-over pattern suggest this likely reflects recovery from Task A belief depression rather than genuine training-induced shifts.

The implication is not that collaboration with AI is harmful, but that mandating a specific synchronous protocol under the infrastructure conditions of this study was associated with worse outcomes than allowing flexible use. Separately, a brief training session that reframed AI as a thought partner was associated with favorable patterns in individual output quality, though the primary continuous specification was null. Belief change was observed but likely reflects carry-over recovery rather than a durable training effect. These are two bounded findings from a single organizational context, not a general verdict on behavioral versus cognitive interventions. Organizations investing in AI adoption should attend to both the structures that shape team collaboration and the mental models that shape individual engagement, while recognizing that the appropriate intervention depends on task demands, infrastructure reliability, and measurement sensitivity.

\subsection*{Ethics Statement}

This study was conducted as part of a scheduled organizational training event at Gap Inc. All participants provided informed consent. This study was not submitted to an Institutional Review Board (IRB) for approval. No deception was involved; both conditions received substantive AI training.

\subsection*{Data Availability}

De-identified survey data and LLM grading scores cannot be shared publicly due to employee confidentiality agreements with Gap Inc. Analysis code is available upon request from the corresponding author. Grading rubrics, survey instruments, and all sensitivity analysis scripts are provided in the appendices and supplementary materials.

\subsection*{Acknowledgments}

We thank Conor Grennan for developing the ``AI Mindset'' partnership training curriculum adapted for the cognitive scaffolding condition. We thank Gap Inc.\ for hosting the AI Learning Day event and supporting the research. We are grateful to our human validation raters for their careful document assessment. All authors are employees of Microsoft Corporation; Microsoft Copilot is a product of Microsoft. The study was designed to evaluate scaffolding interventions for AI use, not to evaluate Microsoft Copilot itself---both conditions had identical Copilot access.

\bibliographystyle{apalike}
\bibliography{references}

\appendix
\renewcommand{\thesection}{\arabic{section}}
\setcounter{section}{7}

\section{Survey Instruments}
\label{app:surveys}

\subsection{Baseline Survey (Pre-Study)}

\emph{Demographics:} Current role/title, job level (IC, Manager, Director+), functional area, tenure.

\emph{AI Experience:} Duration of AI tool usage (never, $<$1 month, 1--2 months, 3--5 months, 6--12 months, $>$1 year), frequency of use (daily, several times/week, weekly, monthly, rarely, never), comfort with Microsoft Copilot (1--5 scale).

\subsection{Post-Task A Survey}

\emph{Task Experience (4 items, 5-point Likert):} Productive and efficient, smooth collaboration flow, confident in output quality, motivated to complete well.

\emph{Copilot Helpfulness (6 items, 5-point Likert + ``Did not use''):} Made work faster, improved quality, supported collaboration, made work more engaging, produced trustworthy outputs, easy to use.

\emph{Behavioral Items (checkbox):} Copilot usage mode (not used, individually, jointly), actions (summarize, generate, edit, brainstorm, learn, get feedback, roleplay).

\subsection{Intermission Survey (Post-Task A, Pre-Task B)}

\emph{AI Beliefs (9 items, 5-point Likert):} AI as Thought Partner (3 items), Exploration \& Experimentation (3 items), Productivity \& Process (3 items). This was the first administration of the belief inventory.

\subsection{Post-Task B Survey}
\label{app:postB}

\emph{Task Experience (5 items, 5-point intensity scale):}

Stem: ``To which extent did you feel\ldots''
\begin{itemize}
  \item \ldots engaged during the task?
  \item \ldots motivated during the task?
  \item \ldots confident in the quality of your final output?
  \item \ldots that the instructions were clear?
  \item \ldots that the workload was manageable?
\end{itemize}

\emph{Copilot Helpfulness (5 items, 5-point intensity scale):}

Stem: ``To which extent did Copilot\ldots''
\begin{itemize}
  \item \ldots help you work faster?
  \item \ldots help you produce higher quality work?
  \item \ldots make the task more engaging?
  \item \ldots feel trustworthy for this task?
  \item \ldots feel easy to use for this task?
\end{itemize}

\emph{Future AI Intent} (3 items, parallel to Post-Task A).

\emph{AI Beliefs} (9 items, repeated; identical to Section~\ref{app:surveys}.3).

\emph{Open-Ended:} What was most valuable about today's training? What would you change?

\section{Document Grading Rubrics}
\label{app:rubrics}

\subsection{Task A Rubric: AI Adoption Action Plan}

\emph{Opportunities Identified (0--6 points):} For each opportunity (max 3, 2 points each): $+$0.5 clear description, $+$0.5 specific noun (named system/stakeholder), $+$0.5 detailed impact, $+$0.5 feasibility assessment.

\emph{Risks Identified (0--6 points):} Parallel structure: description, specificity, impact, testability.

\emph{Action Plan Quality (1--5):} 5 = Executive-ready with timeframes, owners, success criteria; 3 = Steps present but missing details; 1 = Buzzwords only.

\emph{Strategic Insight (1--5):} 5 = Non-obvious insight with downstream effects; 3 = Some organizational awareness; 1 = No strategic thinking.

\subsection{Task B Rubric: Strategic Communications Response}

\emph{Problem Understanding (1--5):} 5 = All three concerns mapped to specific Gap operations; 3 = High-level identification; 1 = Generic response.

\emph{Internal Strategy (1--5):} 5 = Executive-ready with audiences, talking points, actions; 3 = Named audiences; 1 = No useful guidance.

\emph{External Strategy (1--5):} 5 = Constructive, proactive narrative; 3 = Generic ``responsible AI''; 1 = Dismissive.

\emph{Completeness (1--5):} 5 = Fully developed; 3 = Partially developed; 1 = Outline only.

\section{LLM Grading Validation}
\label{app:validation}

\subsection{Grading Protocol}

Total documents graded: 887 (478 Task A, 409 Task B). Final versions: 164 pairs (Task A), 210 individuals (Task B). Primary model: GPT-4o-mini (temperature $= 0$, 3 runs, median score). Validation model: GPT-4o.

\subsection{Determinism}

Table~\ref{tab:determinism} summarizes run-to-run score consistency across the three independent grading passes.

\begin{table}[H]
\centering
\caption{Grading Determinism Across Three Runs}
\label{tab:determinism}
\begin{tabular}{lcc}
\toprule
 & Task A & Task B \\
\midrule
Perfectly deterministic & 53.1\% & 90.7\% \\
Mean score range (non-deterministic) & 0.99 pts & 0.22 pts \\
\bottomrule
\end{tabular}
\end{table}

\subsection{Cross-Model Reliability}

We validated the primary grader (GPT-4o-mini) against a larger model (GPT-4o) on a random sample of 176 documents. Table~\ref{tab:crossmodel} reports agreement metrics.

\begin{table}[H]
\centering
\caption{GPT-4o-mini vs.\ GPT-4o Agreement ($N = 176$)}
\label{tab:crossmodel}
\begin{tabular}{lc}
\toprule
Metric & Value \\
\midrule
Pearson $r$ & 0.922 \\
Spearman $\rho$ & 0.875 \\
ICC(2,1) & 0.921 \\
Mean absolute difference & 1.81 pts \\
Systematic bias (4o-mini $-$ 4o) & $+$0.34 pts \\
\bottomrule
\end{tabular}
\end{table}

\subsection{Word-Count Sensitivity}

Table~\ref{tab:wordcount} reports the correlation between document length and total grading score, along with per-condition mean word counts.

\begin{table}[H]
\centering
\caption{Word-Count Correlation with Quality Scores}
\label{tab:wordcount}
\begin{tabular}{lcc}
\toprule
 & Task A & Task B \\
\midrule
Spearman $\rho$ (score $\sim$ word count) & 0.646 & 0.436 \\
Control mean word count & 740 & --- \\
Treatment mean word count & 454 & --- \\
Word count difference $p$ & .0001 & .017 \\
\bottomrule
\end{tabular}
\end{table}

\subsection{Human Validation Results}

Human validation used 8 trained raters organized in 4 pairs, each pair grading the same 24 documents (61 of 96 assigned documents received complete dual ratings). Raters were blinded to condition and to LLM scores. Tables~\ref{tab:humanagreement}--\ref{tab:rankorder} report the full set of agreement, calibration, and rank-order results.

\emph{Overall Agreement Metrics.} Table~\ref{tab:humanagreement} reports inter-rater reliability among human raters and LLM--human agreement.

\begin{table}[H]
\centering
\caption{Human--Human and LLM--Human Agreement}
\label{tab:humanagreement}
\begin{tabular}{lccc}
\toprule
Metric & Overall & Task A & Task B \\
\midrule
ICC (two-way, agreement) & 0.56 & 0.61 & 0.48 \\
Krippendorff's $\alpha$ & 0.46 & 0.45 & 0.45 \\
Human-Human weighted $\kappa$ & 0.21 & 0.28 & 0.12 \\
LLM-Human weighted $\kappa$ (rater 1) & $-$0.40 & $-$0.41 & $-$0.24 \\
LLM-Human weighted $\kappa$ (rater 2) & $-$0.21 & $-$0.26 & $-$0.14 \\
\bottomrule
\end{tabular}
\end{table}

\emph{Calibration Bias (LLM $-$ Human Average).} Table~\ref{tab:calibration} compares mean LLM scores against mean human averages. The LLM runs systematically generous relative to humans, with larger bias on Task B.

\begin{table}[H]
\centering
\caption{LLM vs.\ Human Score Calibration}
\label{tab:calibration}
\begin{threeparttable}
\begin{tabular}{lccc}
\toprule
Task & LLM Mean & Human Avg Mean & Mean Difference \\
\midrule
Task A & 13.8 & 9.6 & $+$4.2 \\
Task B & 17.3 & 11.8 & $+$5.5 \\
Overall & 15.6 & 10.7 & $+$4.9 \\
\bottomrule
\end{tabular}
\begin{tablenotes}\small
\item \textit{Note:} Mean differences are computed on paired observations and may not equal the difference of rounded column means.
\end{tablenotes}
\end{threeparttable}
\end{table}

\emph{Rank-Order Correlation (Spearman $\rho$, LLM vs.\ Human Average).} Table~\ref{tab:rankorder} reports rank-order agreement at the total-score and per-dimension level. Rank-order preservation is substantially stronger for Task A than for Task B, consistent with the Task B ceiling effect.

\begin{table}[H]
\centering
\caption{LLM--Human Rank-Order Correlation}
\label{tab:rankorder}
\begin{tabular}{lcc}
\toprule
 & Task A & Task B \\
\midrule
Total score & 0.79 & 0.58 \\
Dimension 1 & 0.72 (Opportunities) & 0.54 (Problem Understanding) \\
Dimension 2 & 0.78 (Risks) & 0.49 (Internal Strategy) \\
Dimension 3 & 0.83 (Action Plan) & 0.49 (External Strategy) \\
Dimension 4 & 0.78 (Insight) & 0.51 (Completeness) \\
\bottomrule
\end{tabular}
\end{table}

\emph{Condition Symmetry.} The LLM-human gap did not differ significantly by treatment condition (Task A: Wilcoxon $p = .18$; Task B: $p = .12$).

\section{Sensitivity Analyses}
\label{app:sensitivity}

\subsection{Lee (2009) Trimming Bounds}

Lee bounds address differential attrition by trimming the group with higher response rates from the tails of its outcome distribution, yielding worst-case upper and lower bounds on the ITT effect under a monotonicity assumption.

\begin{table}[H]
\centering
\caption{Lee (2009) Trimming Bounds: All Tasks}
\label{tab:lee-full}
\begin{threeparttable}
\begin{tabular}{llcccc}
\toprule
Task & Bound & Estimate & Bootstrap 95\% CI & Trim \% & Trim Group \\
\midrule
Task A (cont.) & Lower & $-$7.01 & [$-$8.70, $-$5.46] & 23.7\% & Control \\
Task A (cont.) & Point est. & $-$4.96 & --- & --- & --- \\
Task A (cont.) & Upper & $-$3.34 & [$-$5.24, $-$1.62] & 23.7\% & Control \\
\midrule
Task B (cont.) & Lower & $-$1.58 & [$-$2.62, $-$0.58] & 29.3\% & Control \\
Task B (cont.) & Point est. & $+$0.77 & --- & --- & --- \\
Task B (cont.) & Upper & $+$1.90 & [0.54, 3.34] & 29.3\% & Control \\
\midrule
Task B (binary $\geq$20) & Lower & $-$0.10 & [$-$0.25, 0.05] & 29.3\% & Control \\
Task B (binary $\geq$20) & Point est. & $+$0.15 & --- & --- & --- \\
Task B (binary $\geq$20) & Upper & $+$0.31 & [0.16, 0.47] & 29.3\% & Control \\
\bottomrule
\end{tabular}
\begin{tablenotes}\small
\item \textit{Monotonicity assumption:} For Task A, treatment monotonically reduced the probability of producing a document (see Table~\ref{tab:nonproduction}). For Task B, the assumption is less clear-cut; bounds should be interpreted with additional caution.
\end{tablenotes}
\end{threeparttable}
\end{table}

\subsection{Non-Production as Treatment Outcome}

Treatment significantly reduced document production for both tasks, representing an effect on the extensive margin.

\begin{table}[H]
\centering
\caption{Non-Production as Treatment Outcome}
\label{tab:nonproduction}
\begin{tabular}{lccccccc}
\toprule
Task & $N_{\text{Ctrl}}$ & $N_{\text{Trt}}$ & Prod (Ctrl) & Prod (Trt) & OR & 95\% CI & $p$ \\
\midrule
Task A (pairs) & 97 & 97 & 93 & 71 & 0.12 & [0.04, 0.35] & $<$.001 \\
Task B (indiv.) & 194 & 194 & 123 & 87 & 0.47 & [0.31, 0.71] & $<$.001 \\
\bottomrule
\end{tabular}
\end{table}

\subsection{Oster (2019) Coefficient Stability}

Oster bounds quantify how large selection on unobservables would need to be, relative to observed controls, to explain away a coefficient. The $\delta$ column reports the proportional selection required to drive the treatment effect to zero under two assumptions about maximum attainable $R^2$.

\begin{table}[H]
\centering
\caption{Oster (2019) Coefficient Stability Bounds}
\label{tab:oster}
\begin{tabular}{lcccccc}
\toprule
Task & $\beta$ (no ctrl) & $\beta$ (+ wc) & $R^2$ (no ctrl) & $R^2$ (+ wc) & $\delta$ (1.3$\times$) & $\delta$ ($R^2=1$) \\
\midrule
Task A & $-$4.96 & $-$3.34 & 0.173 & 0.370 & \textbf{3.67} & 0.65 \\
Task B & $+$0.77 & $+$0.21 & 0.007 & 0.126 & \textbf{1.17} & --- \\
\bottomrule
\end{tabular}
\end{table}

\subsection{Causal Mediation (Imai et al., 2010)}

The mediation decomposition separates the total treatment effect into the indirect path through word count (ACME) and the direct path (ADE), with bootstrap 95\% confidence intervals. Estimates address how much of the observed treatment effect on quality scores operates through document length rather than content.

\begin{table}[H]
\centering
\caption{Causal Mediation via Word Count}
\label{tab:mediation}
\begin{threeparttable}
\begin{tabular}{llccc}
\toprule
Task & Component & Estimate & 95\% CI & $p$ \\
\midrule
Task A & ACME (via word count) & $-$1.61 & [$-$2.59, $-$0.80] & $< .001$ \\
Task A & ADE (direct) & $-$3.34 & [$-$4.95, $-$1.72] & $< .001$ \\
Task A & Total & $-$4.96 & [$-$6.74, $-$3.33] & $< .001$ \\
Task A & \% Mediated & 32.6\% & [17.7\%, 52.8\%] & --- \\
\midrule
Task B & ACME (via word count) & $+$0.56 & [0.16, 1.05] & .016 \\
Task B & ADE (direct) & $+$0.21 & [$-$0.90, 1.37] & .714 \\
Task B & \% Mediated & 72.9\% & --- & --- \\
\bottomrule
\end{tabular}
\begin{tablenotes}\small
\item \textit{Caveat:} Mediation analysis assumes sequential ignorability---that word count is not itself affected by unmeasured confounders that also affect quality. This assumption is strong and may not hold.
\end{tablenotes}
\end{threeparttable}
\end{table}

\subsection{Session-Effect Sensitivity}
\label{app:session}

The calibration exercise computes the residual Task A treatment effect under a range of assumed pure session effects (in Cohen's $d$), identifying the breakeven point at which a pure session-timing effect would fully account for the observed difference.

\begin{table}[H]
\centering
\caption{Session-Effect Sensitivity Analysis}
\label{tab:session}
\begin{threeparttable}
\begin{tabular}{ccccc}
\toprule
Assumed PM Effect ($d$) & Residual Diff. & Residual $d$ & \% Explained & Significant? \\
\midrule
0.00 (none) & $-$4.96 & $-$0.81 & 0\% & Yes \\
0.10 & $-$4.35 & $-$0.71 & 12.3\% & Yes \\
0.20 & $-$3.74 & $-$0.61 & 24.7\% & Yes \\
0.30 (literature max) & $-$3.13 & $-$0.51 & 37.0\% & Yes \\
0.40 & $-$2.52 & $-$0.41 & 49.4\% & Yes \\
0.50 & $-$1.91 & $-$0.31 & 61.7\% & Yes \\
0.81 (breakeven) & 0.00 & 0.00 & 100\% & No \\
\bottomrule
\end{tabular}
\begin{tablenotes}\small
\item \textit{Note:} $d$ values computed using pooled SD $= 6.11$ from the primary analysis. The observed effect ($|d| = 0.81$) is approximately 3$\times$ larger than the largest circadian decrements in the cognitive psychology literature.
\end{tablenotes}
\end{threeparttable}
\end{table}

\section{Pre-Specified vs.\ Post-Hoc Analytical Decisions}
\label{app:transparency}

\begin{table}[H]
\centering
\caption{Analytical Decision Transparency}
\label{tab:transparency}
\begin{threeparttable}
\small
\begin{tabular}{p{5.5cm}lp{5.5cm}}
\toprule
Decision & Status & Rationale \\
\midrule
Covariates (AI duration, baseline confidence, innovation speed) & Pre-specified & \citet{lin2013}; theoretical relevance \\
Outcome families (Survey: 3; Grading: 2; Belief: 4) & Pre-specified & Grouped by outcome type \\
BH correction within families & Pre-specified & Planned MCP strategy \\
ITT as primary specification & Pre-specified & Standard for RCT \\
HC2 for Task A, CR2 for Task B/Survey & Pre-specified & Unit of analysis \\
Binary threshold for Task B ($\geq$20) & \textbf{Post-hoc} & Based on observed 68.1\% ceiling \\
Compliance group definitions & \textbf{Post-hoc} & Based on observed patterns \\
CLMM for belief change & \textbf{Post-hoc} & Robustness check for ordinal outcomes \\
Lee bounds & \textbf{Post-hoc} & Motivated by observed attrition \\
Oster bounds and mediation & \textbf{Post-hoc} & Motivated by word-count sensitivity \\
Session-effect calibration & \textbf{Post-hoc} & Motivated by AM/PM confound \\
\bottomrule
\end{tabular}
\begin{tablenotes}\small
\item \textit{Note:} This study was not pre-registered. All post-hoc analyses are labeled as such and interpreted accordingly.
\end{tablenotes}
\end{threeparttable}
\end{table}

\section{Partnership Training Curriculum}
\label{app:training}

The cognitive scaffolding condition used the ``AI Mindset'' training curriculum \citep{grennan2023}, a behavioral intervention designed to shift participants' mental models of AI from a transactional frame (AI as search tool or software feature) to a dialogic frame (AI as collaborative thought partner). The training is grounded in the premise that effective AI use is primarily a behavioral challenge rather than a technical one---that is, performance differences stem less from knowledge of system features than from the interaction patterns users adopt.

The curriculum comprised three core components:

\begin{enumerate}
  \item \textbf{Cognitive reframing:} Participants were introduced to behavioral frameworks that challenged default assumptions about AI capabilities and interaction norms, repositioning AI as a collaborator warranting the same iterative engagement one would extend to a human thought partner---including context-setting, follow-up questioning, and real-time course correction.

  \item \textbf{Mental model replacement:} The training explicitly contrasted a ``search engine'' mental model (single-query, extractive) with a ``thought partner'' mental model (multi-turn, generative), using analogies such as the ``smart intern'' metaphor to make the dialogic frame intuitive and actionable.

  \item \textbf{Guided practice:} Participants engaged in structured prompting exercises designed to build fluency with iterative, conversational AI interaction patterns rather than optimizing individual prompt syntax.
\end{enumerate}

The intervention did not introduce any additional system features or technical capabilities beyond what was available to the control condition. The sole differentiator was behavioral: how participants were trained to conceptualize and engage with the AI system. For further information, see \url{https://www.ai-mindset.ai}.

\end{document}